\documentclass[amsmath,amssymb,nofootinbib,showpacs]{JHEP3}
\usepackage{amsmath,amssymb,amsfonts,latexsym,eucal}
\usepackage{graphicx,bm} % Include figure files
\usepackage{dcolumn}     % Align table columns on decimal point

\preprint{UTAP-526, WU-AP/223/05, hep-th/0506019}

\title{On non-uniform smeared black branes}

\author{Hideaki Kudoh
\footnote{Electronic address: kudoh at utap.phys.s.u-tokyo.ac.jp}
\\
Department of Physics, The University of Tokyo, Tokyo 113-0033, Japan   
}

\author{Umpei Miyamoto
\footnote{Electronic address:  umpei at gravity.phys.waseda.ac.jp}
\\
Department of Physics, Waseda University, Okubo 3-4-1, Shinjuku-ku, Tokyo 169-8555, Japan}

\abstract{
We investigate charged dilatonic black $p$-branes smeared on a transverse circle. 
The system can be reduced to neutral vacuum black branes, and we perform static perturbations for the reduced system to construct non-uniform solutions. 
At each order a single master equation is derived, and the Gregory-Laflamme critical wavelength is determined. 
Based on the non-uniform solutions, we discuss thermodynamic properties of this system and argue that in a microcanonical ensemble the non-uniform smeared branes are entropically disfavored even near the extremality, if the spacetime dimension is $D \le 13 +p$, which is the critical dimension for the vacuum case.
However, the critical dimension is not universal. 
In a canonical ensemble the vacuum non-uniform black branes are thermodynamically favorable at $D > 12+p$, whereas the non-uniform smeared branes are favorable at $D > 14+p$ near the extremality.  
}

\maketitle

\begin{document}

%%------------------------------------------------
\section{Introduction }
%%------------------------------------------------

Resolving the structure of vacuum solutions of Kaluza-Klein theory and finding the final fate of the Gregory-Laflamme (GL) instability \cite{Gregory:1993vy,Gregory:1994bj} of black objects has been an intriguing subject in the last decade. 
Recent progress in the subject has revealed many aspects of the problems, but at the same time several important issues remain to be resolved \cite{Kol:2004ww,Harmark:2005pp}.

Perturbative analysis of the vacuum system on a circle has been made extensively, in particular for black holes \cite{Gorbonos:2005px,Karasik:2004ds,Gorbonos:2004uc,Harmark:2003yz} recently. 
For uniform black strings, perturbative non-uniform deformation is allowed only for the critical uniform string that has the static GL mode \cite{Gubser:2001ac}. 
However, the most interesting state of black objects on a circle is in the regime beyond perturbations, and so far only a numerical approach has succeeded. 
The fully nonlinear static solutions with nontrivial horizon geometry have been constructed in 6 dimensions (and, partially, 5 dimensions)
\cite{Wiseman:2002zc,Sorkin:2003ka,Elvang:2004iz,Kudoh:2003ki,Kudoh:2004hs}.
The numerical solutions are constructed by employing the conformal gauge that is a general metric ansatz under axisymmetry and advantageous to numerics. 
The constructed two branches of a vacuum state, black hole and black string, help us to make a phase diagram in this theory~\cite{Kudoh:2004hs}.

To push forward our understanding of vacuum states in many directions, it might be useful to employ a metric ansatz proposed by Harmark and Obers (HO)~\cite{Harmark:2002tr}, which involves just two undetermined functions. 
Although the metric ansatz was originally hypothetical because its proposers could not provide a proof of self-consistency, it has been shown that their simple ansatz is equivalent to the general conformal ansatz by an appropriate coordinate transformation~\cite{Wiseman:2002ti,Harmark:2003eg,Bostock:2004mg}.
A key point is that one can also show the consistency of the associated boundary conditions.

While vacuum solutions represent a rich vacuum structure, charged black objects on a circle are also an interesting subject. A system of charged black objects has in general a thermodynamically stable parameter regime, and related to this fact, there is a conjecture that correlates dynamical stability and thermodynamic stability.
The conjecture is known as the Gubser-Mitra (GM) conjecture \cite{Gubser:2000ec,Gubser:2000mm} (or the correlated stability conjecture \cite{Gubser:2004dr}), which states that for systems with a translational symmetry and an infinite extent the dynamical GL instabilities arise precisely when the system is  thermodynamically unstable. 
There is a lot of evidence for this conjecture, and no fully demonstrated counterexamples have been discovered so far (see, e.g., \cite{Gubser:2004dr,Reall:2001ag,Hirayama:2002hn}). 
Another aspect of introducing charges into the system on a circle is connections with non-gravitational theories~\cite{Aharony:2004ig,Harmark:2004ws}.
Besides, existence of charged non-uniform black branes having higher entropy than their uniform counterparts is conjectured in Ref. \cite{Horowitz:2002ym}.

In this paper, we investigate charged dilatonic black branes smeared on a transverse circle, which we call smeared black branes. 
Following the perturbation theory developed by Gubser \cite{Gubser:2001ac}, we study the stability and possible non-uniform deformation of the smeared black branes. 
To perform the perturbations, we employ the HO metric ansatz, which greatly simplifies the analysis in many respects. 
Based on the solutions of non-uniform smeared black branes, which include vacuum case, we study the thermodynamic properties of the system in detail. 
We think that the perturbative analysis developed in this paper will give a good theoretical basis for constructing fully nonlinear solutions of the HO ansatz, as the original perturbative solution played an important role in constructing nonlinear solutions.

The organization of this paper is as follows. In the next section, we review the smeared black branes in the HO metric ansatz and derive thermodynamical quantities. In Sec. \ref{sec:perturbation}, we perform a perturbative analysis up to the third order. 
The thermodynamics of the smeared black branes is discussed in Sec. \ref{sec:thermodynamics}, and Sec. \ref{sec:conclusion} is devoted to a conclusion.

%%------------------------------------------------
\section{Smeared black  branes}
\label{sec:HO}
%%------------------------------------------------
\subsection{Harmark-Obers ansatz}
\label{sec:HO ansatz}

The part of the classical supergravity action relevant for the considerations in this paper is 
\begin{eqnarray}
I_{D}=\frac{1}{16\pi G_{D}}\int d^{D}x 
\sqrt{-g}\left[R-\frac{1}{2}(\partial \phi)^{2}-\frac{1}{2(p+2)!}e^{\bar{a} \phi}F_{p+2}^{2}\right],
\end{eqnarray}
where $F_{p+2}$ is a $(p+2)$-form field and $\phi$ is a dilaton field. The coupling constant $\bar{a}$ is
$
\bar{a}^{2}=4-{2(p+1)(d-2)}/{(D-2)}
$ with $D=p+d+1$.
From this action, we obtain the following equations: 
\begin{eqnarray}
&&R_{\mu\nu}=\frac{1}{2}\partial _{\mu}\phi\partial _{\nu}
  \phi+\frac{e^{\bar{a}\phi}}{2(p+2)!}
\left[(p+2)F_{\mu}^{\;\;\mu_{2}
\ldots\mu_{p+2}}F_{\nu\mu_{2}\ldots\mu_{p+2}}-\frac{p+1}{D-2}g_{\mu\nu}F_{p+2}^{2}\right],
\label{eq:EinsteinEOM}
\\
&& \Box \phi=\frac{\bar{a}}{2(p+2)!}e^{\bar{a} \phi}F^{2}_{p+2},
\nonumber
\\
&& \partial_{\mu}\left(\sqrt{-g}e^{\bar{a} \phi} F^{\mu\mu_{2}\ldots\mu_{p+2}}\right)=0.
\nonumber
\end{eqnarray}
The form field must satisfy the Bianchi identity,
$ \partial_{[\mu}F_{\mu_{1}\ldots\mu_{p+2}]}=0.$

The ansatz proposed in \cite{Harmark:2002tr} for a charged dilatonic black $p$-brane with transverse space $\mathbb{R}^{d-1}\times S^{1}$ is 
\begin{eqnarray}
&&ds_{D}^{2}=H^{-\frac{d-2}{D-2}}\left[-fdt^{2}+\sum_{i=1}^{p}\left(dx^{i}\right)^{2}+H\left(\frac{L}{2\pi}\right)^{2}\left(\frac{A}{f}dR^{2}+\frac{A}{B^{d-2}}dv^{2} 
  + BR^{2}d\Omega_{d-2}^{2}\right)\right],\nonumber
\\
&&\hspace{2cm} 
f = 1-\left(\frac{R_{0}}{R}\right)^{d-3}, \quad
H = 1+\left(\frac{R_{0}}{R}\right)^{d-3}\!\!\!\!\sinh^{2}\!\alpha,\nonumber
\\
&&
\hspace{2cm}
e^{2\phi}=H^{\bar{a}},
\hspace{1.8cm}
\mathcal{A}_{tx^{1}\cdots x^{p}}=(1-H^{-1})\coth{\alpha}, 
\label{eq:metric}
\end{eqnarray}
where $A=A(R, v)$ and $B=B(R, v)$ are undetermined functions, and $\mathcal{A}$ is the potential of the form field $F=d\mathcal{A}$. 
The constant $\alpha$ is a parameter of electric charge. 
The coordinate $v$ has the periodicity $v=v+2\pi$, and the asymptotic size of the circle is $L$. 
The spacetime has an event horizon at $R=R_{0}$. 
The uniform smeared black $p$-brane is given by setting $A=B=1$.
The consistency between this ansatz and the conformal form for the vacuum case and the charged non-dilatonic case are given in \cite{Wiseman:2002ti,Harmark:2003eg} and \cite{Bostock:2004mg}, respectively. 
The consistency for the most general (charged dilatonic) case is discussed in \cite{Harmark:2003dg} by applying a Lorentz boost and U-duality.

The remarkable character of this metric is that the EOMs for the two unknown functions $A$ and $B$ are independent of the value of $p$ and the charge parameter $\alpha$. 
Thus the EOMs are the same as those of the neutral non-dilatonic black branes on ${\mathbb R}^{d-1} \times S^1$, which we represent in Appendix \ref{sec:vacuum}. Consequently, the problem of finding solutions of charged non-uniform smeared branes is mapped to the problem of neutral non-uniform strings with the effective spacetime dimension $\mathcal{D}$,~\footnote{
Generating (near-extremal) charged solutions from uncharged solutions via the M-theory lift-boost-reduction is possible. This kind of solution-generating technique was used in \cite{Aharony:2004ig,Harmark:2004ws,Hassan:1991mq}.
}
\begin{eqnarray}
\mathcal{D} \equiv D-p=d+1. 
\end{eqnarray}
For example, the non-dilatonic charged cases are $(D, p, d; \bar{a}) =$ (11, 5, 5; 0) and (10, 3, 6; 0), and their effective dimensions are $\mathcal{D}=6$ and $7$, respectively. The two cases correspond to M5- and D3-branes.
In this paper, we take $\mathcal{D}$ as a parameter, and we study perturbations in a wide range of $\mathcal{D}$.

%%------------------------------------------------
\subsection{Thermodynamic quantities}
\label{subsec:thermodynamic q}
%%------------------------------------------------

Thermodynamic quantities of the smeared black branes are calculated from the metric (\ref{eq:metric}). The mass, temperature, entropy and charge are given by
\begin{eqnarray}
M
&=&  {\mathcal G}  \left( \frac{L R_0}{2\pi}\right)^{d-3} (d-3) 
\left[
\frac{d-2}{d-3}  -\chi+ \sinh ^2 \alpha
\right],\nonumber
\\
T&=&\frac{d-3}{2R_0 L \sqrt{A_h} \cosh\alpha   },\nonumber
\\
S
&=&
    4\pi {\mathcal G}  \sqrt{A_h} 
   \left( \frac{L R_0}{2\pi}\right)^{d-2}
   \cosh \alpha,\nonumber
\\
Q
&=&
    {\mathcal G} 
\left( \frac{L R_0}{2\pi}\right)^{d-3}(d-3)\cosh\alpha\sinh\alpha,
\label{eq:M,T,S,Q}
\end{eqnarray}
where we have used ${\mathcal G} \equiv L V_p \Omega_{d-2} /16\pi G_D$. 
$A_{h}(R_0) \equiv A(R_{0}, v)$ and $\chi= \chi(R_0)$ are constants for a given horizon radius.
The latter is defined by the asymptotic behavior of the metric functions:
\begin{eqnarray}
A, B \simeq 1-\chi\left(\frac{R_{0}}{R}\right)^{d-3},
\quad
 \mbox{for}\;\; R \gg R_{0}.
\label{eq:def chi}
\end{eqnarray}
$V_{p}$ is the volume of $x^{i}$ ($i=1, 2, \cdots, p$) directions. 
It is not necessary to take the coordinate $x^{i}$ to be periodic, but we do it formally so that we may present a finite expression of mass and charge.
The volume of the unit $n$-dimensional sphere is $\Omega_{n}= 2\pi^{(n+1)/2}/\Gamma\left[(n+2)/2\right]$. 
For a regular event horizon, $A_{h}$ is independent of $v$ so that the zeroth law of thermodynamics holds. 
The tensions, which are asymptotic charges accompanied by the $S^{1}$ compactification, are given by 
\cite{Harmark:2003dg,Kol:2003if,Townsend:2001rg,Traschen:2001pb}
\begin{eqnarray}
    L \mathbb{T}
&=& 
    {\mathcal G}     \left( \frac{L R_0}{2\pi} \right)^{d-3}
    \left[ 1 - (d-3) (d-2) \chi \right], \nonumber
\\
    L_i {\mathbb{T}}_i
&=& 
    {\mathcal G}    
    \left( \frac{L R_0}{2\pi} \right)^{d-3}
    \left[
        1 + (d-3) ( \sinh^2 \alpha - \chi )
    \right],
\end{eqnarray}
where $\mathbb{T}$ and ${\mathbb{T}}_i$ are the tensions along the $v$-direction and along the $i$-th world-volume with circle length $L_i$, respectively.  
Defining the dimensionless tension $n$ by  
\begin{eqnarray}
n \equiv \frac{ L {\mathbb{T}}  }{M-M_{el}},
\label{eq:def n}
\end{eqnarray}
where $M_{el} \equiv \left[ (d-1) L_i {\mathbb{T}}_i - L {\mathbb{T}} -M  \right]/(d-2)$ is ``electric"  mass \cite{Harmark:2004ws}, the Smarr formula then takes the form
\begin{eqnarray}
 T S = \frac{d-2-n}{d-1} (M- M_{el}). 
\label{eq:smarr}
\end{eqnarray}
Note that for the neutral case ($\alpha=0$), we have $M_{el}=0$ and that then the dimensionless tension becomes the one commonly employed for vacuum black branes.

The specific heat $C_{Q}$ and the isothermal permittivity $\epsilon_{T} $ are computed as
\begin{eqnarray}
C_{Q} &\equiv&
 \left(  \frac{\partial M}{\partial T}\right)_Q
= \frac{V_{p} \Omega_{d-2}L^{d-1} }{4 G} \left( \frac{R_0}{2\pi} \right)^{d-2}
\frac{  \left[d-3 +(d-1)\cosh 2\alpha\right]\cosh \alpha }{-(d-3)+ (d-5) \cosh 2\alpha},
\\
\epsilon_{T} 
&\equiv&
 \left(  \frac{\partial Q}{\partial \Phi }\right)_T
= 
\frac{ V_{p} \Omega_{d-2}L^{d-2} }{32\pi G}
  \left( \frac{R_0}{2\pi} \right)^{d-3}(d-3)
 \left[(d-3) - (d-5) \cosh 2\alpha\right]\cosh^2 \alpha,
\nonumber
\end{eqnarray}
where 
\begin{eqnarray}
\Phi  = \tanh \alpha 
\end{eqnarray}
is an electric potential energy at the horizon. 
The specific heat changes the sign from minus to plus at a critical value of $\alpha$, if $d\geq 6$. For a fixed charge $Q$, this implies that 
a thermodynamically stable region exists for the smeared black branes in the parameter region in which black branes suffer from dynamical GL instability \cite{Bostock:2004mg}.  
However, we should consider the thermodynamic ensemble that allows the charge to change. In other words, we should also take into account the isothermal permittivity, which probes the thermodynamic stability under changes of the charge~\cite{Hirayama:2002hn,Ross:2005vh}.  
One finds that the specific heat and isothermal permittivity always have opposite signs from the above expression. Consequently, the two conditions $C_{Q}>0$ and $\epsilon_{T}>0$, which guarantee thermodynamic stability in the grand canonical ensemble, cannot hold simultaneously. 
That is to say, the smeared black brane background is never thermodynamically stable. 
This is perfectly compatible with the existence of a static mode that is a signal of the onset of unstable modes. 
We discuss the static mode in detail in the next section.\footnote{
A partial preliminary analysis of dynamical s-wave perturbations is found in \cite{Kang:2005is}.
}

%%------------------------------------------------
\section{Static perturbation}
\label{sec:perturbation}
%%------------------------------------------------

According to the perturbation theory in Ref.~\cite{Gubser:2001ac}, we perform static perturbations of the smeared black $p$-branes.\footnote{
Readers who are interested in the details in this section can examine a Mathematica notebook~\cite{Mathematica}
} In general, the static perturbations are easier to perform than dynamical perturbations, and they allows us to construct a non-uniform solution as well as to determine the GL critical waves.  
For convenience, we transform the metric functions and the coordinates as
\begin{eqnarray}
&& A=e^{a},\;\;\;B=e^{b},
\nonumber\\
&&
 x = \frac{L}{2\pi} v , \quad
 y = \frac{L}{2\pi} R. 
\end{eqnarray}
It is possible to rescale the entire metric so that the event horizon can be located at $y=1$ in the new coordinates. 
In the following analysis we adopt this normalization.

We expand the metric function $X(x, y)$ ($X=a, b$) around the uniform solution as follows:
\begin{eqnarray}
X (x,y) = \sum_{n=0}^{\infty} \epsilon^n X_n(y) \cos n K x.
\label{eq:Def Expansion}
\end{eqnarray}
Here, $\epsilon$ is a small parameter of expansion.
The function $X_{n}$ and wave number $K=2\pi/L$ are also expanded as 
\begin{eqnarray}
X_n(y)=\sum_{p=0}^{\infty} \epsilon^{2p} X_{n,p}(y),\;\;\;
K =\sum_{q=0}^{\infty}\epsilon^{2q} k_{q},
\end{eqnarray}
where $X_{0, 0}(y)=0$ is imposed. 
Some leading order terms are
\begin{eqnarray}
 X(x, y)
&  = &
  \epsilon X_{1,0}(y)\cos k_{0}x+\epsilon^2 \left[X_{0, 1}(y)+X_{2, 0}(y)\cos 2k_{0}x\right]\nonumber
\\
&& 
+ \epsilon^{3}\left[X_{1, 1}(y)\cos k_{0}x-X_{1, 0}(y)k_{1}x\sin k_{0}x+X_{3, 0}(y)\cos 3k_{0}x\right]+O(\epsilon^4).
\label{eq:expansion}
\end{eqnarray}

Having expanded the metric functions as above, all we have to do is to solve the ordinary differential equations (ODEs) for $X_{n, p}(y)$ at each order of $\epsilon$. 
Before entering into detail, however, it would be suitable to outline our calculations.
First of all we should notice that in this perturbation theory the asymptotic circle length (or the wave number) is not fixed at all, and we have to determine it by solving the perturbations. 
At the first order $O(\epsilon)$, we look for a solution $X_{1, 0}$, which corresponds to the GL static mode. Thus, we can show the instability of the uniform solution by finding the GL critical wave number $k_{0}$. 
At the second order $O(\epsilon^{2})$, a back-reaction leads to nontrivial $X_{0, 1}$ and $X_{2, 0}$. 
One of them is the Kaluza-Klein (KK) massive mode, which falls off exponentially for a large $y$. The other is a massless mode, which determines the mass of smeared branes as an asymptotic charge. 
Therefore, one might consider the second-order calculation to be enough to see a thermodynamical symptom of the non-uniform deformation.
It is known, however, that to calculate entropy difference between non-uniform and uniform solutions for the same mass one needs third-order perturbations, at least in the uncharged case. 
This is also the case for our general situation.

It is important to note the advantage of working in the HO metric. 
The metric ansatz reduces the number of equations of motion (EOMs) greatly. 
As discussed in the original paper \cite{Harmark:2002tr}, the EOMs do not contain the charge parameter $\alpha$ and the spatial world-volume $p$. 
The EOMs for most general charged diatonic cases are equivalent to those for neutral black strings on a cylinder $\mathbb{R}^{\mathcal{D}-2}\times S^{1}$ with $\mathcal{D}=D-p=d+1$. That is, we do not need to perturb the form and dilation fields at any order.

It is worthwhile to consider the meaning of the parameter $\epsilon$.
In a leading order, this parameter $\epsilon$ can be related to the parameter used in the literature \cite{Gubser:2001ac},
\begin{eqnarray}
\lambda \equiv \frac{1}{2}\left(\frac{R_{max}}{R_{min}}-1\right). 
\label{eq:lambda}
\end{eqnarray}
Here, $R_{max}$ and $R_{min}$ are the maximum and minimum of the proper horizon radius of the black object so that $\lambda$ means geometrically the non-uniformity of the black brane. 
Substituting the expansion (\ref{eq:Def Expansion}) into the metric (\ref{eq:metric}), the parameter $\epsilon$ is identical to $\lambda$ at the linear order if we set $b_{1,0}(1)=2$, which is possible due to the linearity of the differential equation at this order. 
Consequently, the expansion parameter $\epsilon$ describes in the above sense how the black brane is non-uniform.  
However, the relation between $\epsilon$ and $\lambda$ becomes nonlinear when we take into account higher-order perturbations.   
Because of the nonlinear corrections, the naive identification between $\epsilon$ and $\lambda$ is apparently invalid in the regime.  
After all, $\epsilon$ is just an expansion parameter and is no more and no less than the parameter controlling the amplitude of the perturbations.

%%-----------------------------------------------
\subsection{First-order perturbation $X_{1,0}$}
\label{seq:first}
%%-----------------------------------------------

Let us begin with the first-order perturbation. 
Substituting the expansion (\ref{eq:expansion}) into the Einstein equation (\ref{eq:EinsteinEOM}), we obtain a master equation for $b_{1, 0}$, which determines the GL critical wave number $k_0$,
\begin{eqnarray}
    {\mathcal L}_{1} b_{1, 0}=0,
    \label{eq:first-master}
\end{eqnarray}
where
\begin{eqnarray}
&&{\mathcal L}_1=\frac{d^{2}}{dy^{2}} + U \frac{d}{dy} + V_{(21)},
\nonumber
\\
&& U(y)= \frac{(d-1)f^2 - (d-3)(3d - 8)f  + (d-3)^2 }{yf[d-3+(d-1)f]},
\nonumber
\\
&& V_{(ij)}(y)= -  \frac{i (d-3)^2(f-1) +j k_{0} ^2y^2 [d-3 + (d-1)f]}
            {y^{2}f[d-3+(d-1)f]}.
\end{eqnarray}
After some manipulation, one recognizes that the problem is the same as solving the Schr\"{o}dinger-type equation with energy $-k_0^2$, although the potential is rather complicated in the present case.\footnote{
Another type of single linear second-order ODE was discussed in \cite{Kol:2004pn}, based on perturbations of Euclidean Schwarzschild black holes.
However, since the equation has a singular point between the horizon and the infinity, it is clearly less tractable compared to our master equation.
}

One can solve algebraically for $a_{1, 0}$, with the result 
\begin{eqnarray}
a_{1, 0}(y) = 
\frac{2(d-2)f(b_{1, 0} +y b_{1, 0}')}{d-3 + (d-1)f} ,
\label{eq:a10}
\end{eqnarray}
where the prime denotes the derivative with respect to $y$.

The horizon is a regular singular point of the differential equation, and demanding the regularity of the perturbations on the horizon, an appropriate boundary condition is required at the horizon:
\begin{eqnarray}
b_{1, 0}'(1) + \left( 2 - \frac{k_0^2}{d-3} \right) b_{1,0}(1)=0.
\label{eq: bc for b_10}
\end{eqnarray}
According to the discussion in the beginning of this section, we take 
\begin{eqnarray}
b_{1,0}(1)=2. 
\end{eqnarray}
From the zeroth law of thermodynamics, all Kaluza-Klein modes of $a(x,y)$ must vanish at the horizon. So we have $a_{1, 0}(1)=0$, as is also evident from (\ref{eq:a10}) and the finiteness of $b_{1,0}(1)$.

\begin{figure}[tbp]  %%[thbp]
\begin{center}
\includegraphics[width=7cm]{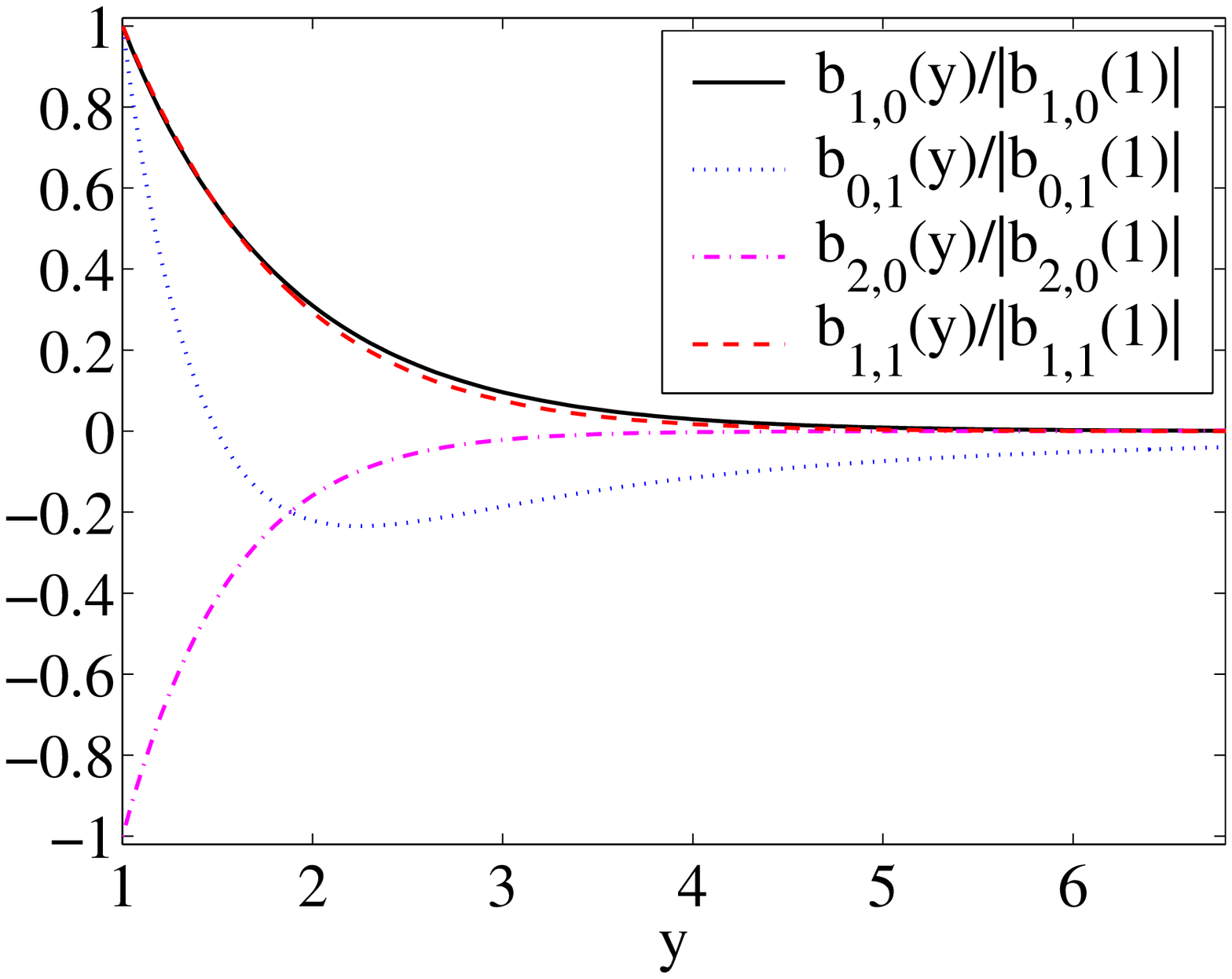}
\qquad
\includegraphics[width=7cm]{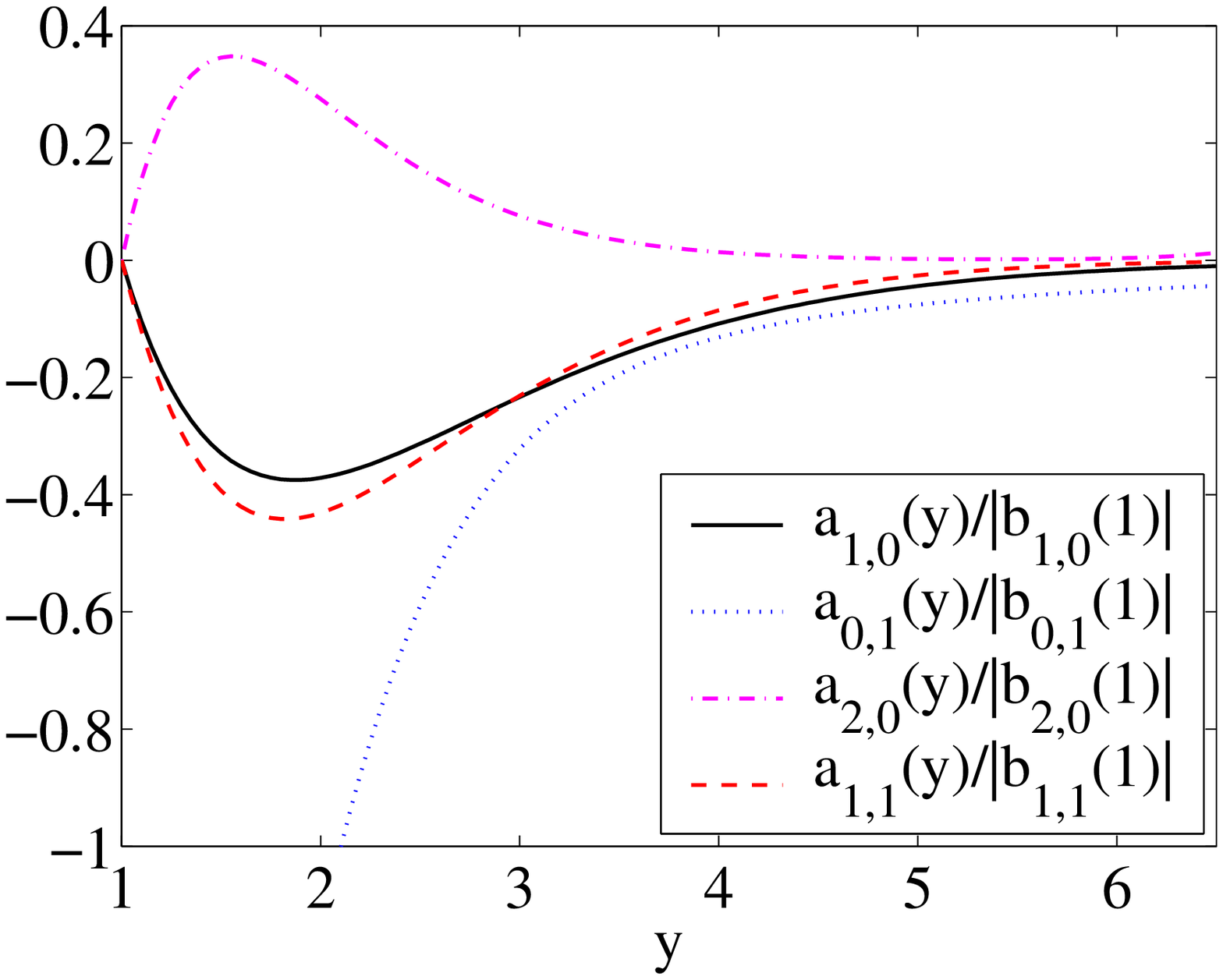}
\caption{
\label{fig:perturbations}
The numerical plots of metric perturbations for $\mathcal{D}=6$. The event horizon is located at $y=1$. 
Each line is normalized by the absolute value of $b$ at the horizon. 
Only $b_{0,1}$ (and $a_{0,1}$) is the zero mode, so it shows slow decay at the asymptotics, compared to other KK massive modes.
}
\end{center}
\end{figure}

The perturbation equation has two independent solutions, and at infinity the respective solutions are exponentially growing or decaying. The asymptotic flatness implies the growing mode is absent at infinity.
The asymptotic behavior of the decaying mode is 
\begin{eqnarray}
  b_{1,0} \sim e^{-k_0 y} y^{(d-4)/{2}}. 
\end{eqnarray}
Thus, finding the solution of Eq. (\ref{eq:first-master}) is a one parameter shooting problem, which is easy to carry out. 
The GL critical wave number $k_0$ is determined for several dimensions and is summarized in Table \ref{table:k0}.\footnote{
Note that we have performed the calculations up to $\mathcal D=20$, although we show only a part of them in the Tables.
\label{footnote:D=20}
}
The values of $k_0$ do of course accord with previous results \cite{Gregory:1993vy,Gubser:2001ac,Wiseman:2002zc,Kol:2004pn}. 
The numerical plots of the solutions are given in Fig. \ref{fig:perturbations}.

%%%%%%%%%%%%%%%%%%% Table %%%%%%%%%%%%%%%%%%%%%%%%%%%%%%%%%%%%
\begin{table}[tbp]
\begin{center}
\begin{tabular}{rc||c|c|c|c|c|c|c|c|c|c|c }
\hline \hline
& ${\mathcal{D}}=d+1$ & 5 & 6 & 7 & 8 & 9 & 10 & 11 & 12 & 13 & 14 & 15
\\ \hline  
&$k_{0} $ &0.876&1.27&1.58& 1.85&2.09& 2.30& 2.50& 2.69& 2.87 & 3.03 & 3.19
\\ \hline  
&$k_{1} $ &1.62& 2.20&2.71& 3.09&3.29& 3.23& 2.91& 2.29 & 1.30& -0.05 & -1.83
\\
\hline \hline
\end{tabular}
\caption[short]{
Gregory-Laflamme critical wave number $k_0$ for the effective dimension 
${\mathcal{D}=5 \sim 15}$. 
$k_1$ is the next-order correction to the critical wave number. 
The horizon radius of the uniform solution is $y=1$. 
An approximate algebraic equation of the GL critical wave number for ${\mathcal{D} \lesssim 9}$ is given in Eq. (\ref{eq:approximate GL eq}). 
}
\label{table:k0}
\end{center}
\end{table}
%%%%%%%%%%%%%%%%%%% Table %%%%%%%%%%%%%%%%%%%%%%%%%%%%%%%%%%%%%

Let us take a closer look at the solution near the horizon. 
We assume the following series expansion for $b_{1,0}$,
\begin{eqnarray}
    b_{1,0}  = e^{-k_0 (y-1)} \sum_{n=0}^\infty \beta_n (y-1)^n  .
\end{eqnarray}
Substituting this into Eq. (\ref{eq:first-master}), we obtain a five-term recurrence equation. The recurrence equation can be solved by employing the continued fraction method, which determines the critical wavelength. However, the shooting method is more tractable than the fraction method, and thus we refrain from presenting the latter. 
The first few terms are still useful to observing the near horizon behavior:
\begin{eqnarray}
&&\beta_0=2,\nonumber
\\  
&&\beta_1=2\left( k_{0}-2 + \frac{k_{0}^{2}}{d-4}\right),\nonumber
\\  
&&\beta_2= 8 -2d + \frac{k_{0}}{2}
    \left[  \frac{k_{0}^{2}(k_{0}+4d-12)}{(d-3)^2}  -8 + 6k_{0}  \right],\nonumber
\\  
&&\beta_3 = \cdots.
\end{eqnarray}
If we truncate the series expansion at some level and match the truncated solution to the asymptotic solution at an appropriate radius, we might be able to obtain an approximate solution of the critical wavelength. 
For the truncation at $O(\beta_3)$ and matching at $y\approx 2$, we find a fourth-order algebraic equation, 
\begin{multline}
k_{0}^4 + (d-3)\\
\times\left[
4 k_{0}^3 
+ 2 \frac{ \left\{138 + (1 + d) \left[-49 + 3 (1 + d)\right]\right\}k_{0}^2}{d-12}
- \frac{ 4(d-16)(d-3) k_{0} }{d-12}
- 4 (d-3)^2
\right] \approx 0.
\label{eq:approximate GL eq}
\end{multline}
A real solution of this algebraic equation provides a good approximate solution of the GL critical wave for $\mathcal{D}\lesssim 9$ with small errors ({a few percent}).
A more accurate and readily available algebraic equation could be obtained by pushing the truncation level up to $O(\beta_5)$. 
But the algebraic equation becomes eighth order and it has no analytic solution in general.

We have seen the existence of the GL critical wave number $k_{0}$ for an arbitrary value of $\alpha$, which does not appear in the master equation (\ref{eq:first-master}). In other words, the smeared black branes suffer from the GL instability, irrespective of their charge. As mentioned in Sec. \ref{subsec:thermodynamic q}, however, the non-extremal smeared black brane cannot be locally thermodynamically stable for any value of $\alpha$. Therefore, the existence of the GL critical wave number does not imply the breakdown of the GM conjecture.

%%%%%%%%%%%%%%%%%%%%%%%  Fig.  %%%%%%%%%%%%%%%%%%%%%%%%%%%%%%%
%%%%%%%%%%%%%%%%%%%%%%%%%%%%%%%%%%%%%%%%%%%%%%%%%%%%%%%%%%%%%%

%%%----------------------------------------------
\subsection{Second-order perturbation }
\label{sec:second}
%%%----------------------------------------------

The first-order perturbations determine the GL critical wave number $k_0$, and they are localized near the horizon due to the rapid decay at the far region.
The localized modes produce sources for higher-order perturbations that give back-reaction to the background geometry. 
At the second order $O(\epsilon^2)$, two types of mode appear: the KK modes, which decay exponentially, and the zero modes, which have power-law decay at the asymptotics. 
Since the power-law decay of the zero mode gives finite contributions to the asymptotic charges, the zero modes have rather different characters compared with the KK modes.

%%%%%%%%%%%%%%%%%%%%%%%%%%%%%%%%%%%%%%%%%%%%%%%%%%%%%%%%%%%%%%%%%%
\subsubsection{Zero modes $X_{0,1}$}
\label{sec:second-zero}
%%%%%%%%%%%%%%%%%%%%%%%%%%%%%%%%%%%%%%%%%%%%%%%%%%%%%%%%%%%%%%%%%%

The equations for the zero modes are rather complicated. 
The ODEs for $a_{0, 1}$ and $b_{0,1}$ are
\begin{eqnarray}
&&b_{0, 1}'' + \frac{3y^2-1}{y(y^2-1)}  b_{0, 1}'
- \frac{4}{y^2 -1} (a_{0, 1} - b_{0, 1})  = S_{b_{0, 1}}(b_{1, 0}; k_{0}),\label{eq:eqforb0}\nonumber\\
&&a_{0, 1}'' + \frac{3y^2-1}{y(y^2-1)} a_{0, 1}'
-\frac{12}{y^2 -1} (a_{0, 1} - b_{0, 1})=S_{a_{0, 1}}(b_{1, 0}; k_{0}),\label{eq:eqfora0}
\end{eqnarray}
where we abbreviate the source term to $S_{X_{0,1}}$ which consists of a quadratic expression in $b_{1, 0}$ and its first and second derivatives. 
All second derivatives are eliminated from the source term using the equation of the first-order perturbations. 
The critical wave number appears in the source term.

To have decoupled equations, it is convenient to introduce new variables,
\begin{eqnarray}
c_{0, 1}\equiv b_{0, 1}-a_{0, 1},\;\;\;
d_{0, 1}\equiv b_{0, 1}+a_{0, 1}.
\end{eqnarray}
Then the equations are 
\begin{eqnarray}
c_{0, 1}'' + \frac{ f+d-3 }{ y f} c_{0, 1}'
- \frac{2(d-3)^2}{y^2 f} c_{0, 1} 
&=& S_{c_{0, 1}}(b_{1, 0}; k_{0}),\nonumber\\
d_{0, 1}'' + \frac{ f+d-3 }{ y f} d_{0, 1}' + \frac{2(d-3)(d-1)}{y^2 f}c_{0, 1} 
&=& S_{d_{0, 1}}( b_{1, 0}; k_{0}),
\label{eq:c,d_{0, 1}}
\end{eqnarray}
where 
\begin{eqnarray}
\hspace{-2cm}&&S_{c_{0, 1}}(b_{1, 0}; k_{0})
=
- 
\frac{1}{4y^2f}
\left\{
 2(d-3)^2 (a_{1,0}-b_{1,0})^2 + (d-2) k_0^2 y^2 b_{1,0} 
 \left[2a_{1,0} +(d-3)b_{1,0}\right]
\right\},
\nonumber\\
&&S_{d_{0, 1}}(b_{1, 0}; k_{0})
=
\frac{1}{4y^2f}
\left\{
 2(d-3)(d-1) (a_{1,0}-b_{1,0})^2 
+ (d-2) k_0^2 y^2 b_{1,0} 
 \left[2a_{1,0} +(d-3)b_{1,0}\right]
\right\}.
\nonumber\\
\end{eqnarray}
Note that from the Einstein equations we also have an additional equation which takes schematically the form $d_{0,1}'=F(b_{0,1},c_{0,1},b'_{0,1},c'_{0,1}; k_0)$. 
It is not an independent equation; the second-order equation (\ref{eq:c,d_{0, 1}}) for $d_{0,1}$ follows from the first-order equations.

If we assume the regularity of the perturbed quantities, 
we obtain the horizon boundary conditions by setting $y=1$:
\begin{eqnarray}
&&c_{0, 1}'(1) = 2(d-3)c_{0, 1}  - \frac{1}{4}\left[ 2(d-3)+ (d-2)k_0^2\right] b_{1, 0}^2,\nonumber\\
&&d_{0, 1}'(1) = - 2 (d-1)c_{0,1} + \frac{1}{4}\left[ 2(d-1)+(d-2)k_0^2) \right]b_{1,0}^2.
\label{eq:BC-d0}
\end{eqnarray}
Since at the asymptotic region the source terms decay as $O(e^{-2k_{0}y})$, we can neglect the right-hand side of (\ref{eq:c,d_{0, 1}}). 
Then the leading asymptotic behavior of $d_{0,1}$ and $c_{0,1}$ are 
$d_{0,1} \sim 1/y^{(d-3)}$ and $c_{0,1} \sim 1/y^{2(d-3)} $. 
Interestingly, the power-law decay of $c_{0,1}$ is so rapid that it does not contribute to the asymptotic charges, due to Eq. (\ref{eq:def chi}).

Now let us discuss how we can solve these equations.
The shooting parameter of this system is $c_{0,1}(1)$. The value of $d_{0,1}(1)$ at the horizon is not a shooting parameter. It is an arbitrary constant. 
This is because $d_{0,1}$ has a shift symmetry in the differential equation and the boundary condition, 
\begin{eqnarray}
    d_{0,1} \to d_{0,1} + \mathrm{const.}
\label{eq:shift}
\end{eqnarray}
Clearly, we cannot choose the constant in Eq.~(\ref{eq:shift}) arbitrarily: If we shift the value of $d_{0,1}$, the value of $a_{0,1}$ on the horizon changes, which means that the temperature and entropy of the black brane changes. 
The constant should be fixed by imposing the asymptotic flatness on $d_{0,1}$, after one obtains a solution.

The numerical plot of the solutions are presented in Fig. \ref{fig:perturbations}, in which one can see the power-law decay of the solution. 
If we expand the constant $\chi$ in Eq. (\ref{eq:def chi}) as
\begin{eqnarray}
    \chi = \chi_1 \epsilon^2 + O(\epsilon^4),
\end{eqnarray}
the asymptotic behaviors of $a_{0,1}$ and $b_{0,1}$ are obtained from $d_{0,1}$ as 
\begin{eqnarray}
 a_{0,1}, b_{0,1} \sim -  \frac{\chi_1}{y^{d-3}}. 
\label{eq:dumping}
\end{eqnarray}
It is worthwhile to note that from the first law of the thermodynamics the following relation holes,
\begin{eqnarray}
  \chi_1 = - \frac{a_{0,1}(1)}{2}. 
\label{eq:chi_1 and a_01}
\end{eqnarray}
The results of numerics for various effective dimensions $\mathcal{D}$ are shown in Table~\ref{table:perturbations}.

%%%%%%%%%%%%%%%%%%%%%%%%%%%%%%%%%%%%%%%%%%%%%%%%%%%%%%%%%%%%%%%%%%
\subsubsection{KK modes $X_{2,0}$ }
\label{sec:second-kk}
%%%%%%%%%%%%%%%%%%%%%%%%%%%%%%%%%%%%%%%%%%%%%%%%%%%%%%%%%%%%%%%%%%
The KK modes at this order are $a_{2, 0}$ and $b_{2, 0}$. 
The function $a_{2, 0}$ can be solved algebraically, 
\begin{multline}
a_{2,0}=
 \frac{2f (d-2) ( y b_{2,0})' }{d-3 + ( d-1) f}
- \frac{ a_{1,0} \left[ a_{1,0} - (d-1) b_{1,0}\right]}{8}
- \frac{ (d-2) f b_{1,0} }{4\left[ d-3+(d -1)f\right]^2}\\
\times \Big(
  2 (d-3) (d-2) a_{1,0}
+ b_{1,0}\left\{ 2(d-2)k_{0}^2 y^2 -(d-3)^2  + f \left[5+(d-4)d \right] \right\}
\Big).
\end{multline}
Here the source term consists of $b_{1, 0}, b_{2,0}$ (and also $a_{1, 0}$). 
A master equation for $b_{2, 0}$ is similar to the first-order perturbation:
\begin{eqnarray}
b''_{2,0}
+ U b'_{2,0}
+ V_{(12)} b_{2,0} = S_{b_{2,0}} (b_{1,0};k_0),
\end{eqnarray}
where the source term $S_{b_{2,0}}$ is quadratic in $b_{1,0}$ and its first derivative. The explicit form is
\begin{multline}
S_{b_{2,0}} 
=
\frac{d-3}{4y^2f}
\Biggl[
 a_{1,0}^2  
+  a_{1,0}b_{1,0} \biggl\{ 
         (d-5) - 
         \frac{4f (d-3)(d-2)^2 }{\left[d-3+ (d-1)f\right]^2} 
       \biggr\}- \frac{ 2b_{1,0}^2 }{\left[d-3 + (d-1)f\right]^2 }
\\
\times\biggl\{
{(d-3) } \left[ 3 - d - 2(d-2) k_0^2 y^2   \right]
      - f \left\{ d [ 13-6d+d^2+2(d-2)k_0^2 y^2]  -12  \right\}      
\\
  + \frac{f^2}{d-3}  \left[ 33 - 56 d + 36 d^2 - 10 d^3 + d^4 - 2 (d-2)  
           { ( d-1  ) }^2 k_0^2 y^2 \right] 
\biggr\} 
\Biggr],
\label{eq:Sb20}
\end{multline}
where $a_{1,0}$ in the source term can be eliminated by Eq. (\ref{eq:a10}).

The asymptotic behavior of these KK modes are $b_{2,0}, a_{2,0}\sim \exp(-2k_{0}y)$, and the horizon boundary conditions are 
\begin{eqnarray}
b_{2,0}'(1) + \left( 2 - \frac{k_0^2}{d-3}\right) b_{2,0}
 = 
\left[ \frac{1}{2}+ \frac{(d-2)k^2_0}{ d-3 }  \right]~ b_{1, 0}^2 
\end{eqnarray}
and $a_{2,0}(1) =0$.
Since the critical wave number $k_0$ is given at the first order, this ODE is a shooting problem with one shooting parameter, i.e., amplitude at the horizon $b_{2, 0}(1)$. 
Applying the same numerics as for the first order, we can easily obtain the solution.  
The parameter that gives a regular solution is summarized in Table \ref{table:perturbations}.

%%%%%%%%%%%%%%%%%%% Table %%%%%%%%%%%%%%%%%%%%%%%%%%%%%%%%%%%%%
\begin{table}[t]
\begin{center}
\begin{tabular}{rc||c|c|c|c|c|c|c|c|c|c|c }
\hline \hline
&${\mathcal{D}}$ & 5 & 6 & 7 & 8 & 9 & 10 & 11 & 12 & 13 & 14 & 15
\\ 
\hline  
&$a_{0,1}(1)$&-1.63&-1.30& -1.20 & -1.15&-1.13& -1.11&-1.10&-1.09& -1.08&-1.07 & -1.01
\\
\hline
&$b_{0,1}(1)$& -0.65&~0.36&~1.05&~1.65&~2.21&~2.75&~3.29&~3.8&~4.3&~4.9&~5.4
\\
\hline
&$b_{2,0}(1)$& -2.83&-3.96 &-5.09&-6.22&-7.36&-8.51&-9.66&-11&-12&-13& -14
\\
\hline
&$b_{1,1}(1)$&~19.7&~21.9&~20.2&9.91&-1.40&-8.25&-17.4&-27&-25&-60&-69
\\ \hline \hline
&$\mu_{1}$   &~1.45&~3.03 &~4.69&~7.30&~7.41&~7.95&~7.66&~6.3&~3.6&-0.67&-6.8
\\  \hline
& $s_{1}$ &~2.89  &~4.55 &~6.25 &~7.78 &~8.90&~9.27&~8.76&~7.1&~4.0&-0.74&-7.4
\\ \hline 
&$\tau_{1}$&-1.04&-1.08&-1.11& -1.09&-1.01&-0.85& -0.62 & -0.31 &~0.09&~0.56&~1.1
\\ 
\hline \hline
\end{tabular}
\caption[short]{
Summary of numerical results for the static perturbations. 
The range of effective spacetime dimensions is ${\mathcal D}=D-p =5 \sim 15$. The amplitude of the linear perturbation is set as $b_{1,0}(1)=2$. 
The quantities $\mu_1$ and $s_1$ change their signs above $\mathcal D =13$, whereas $\tau_1$ changes its sign above $\mathcal D =12$. 
}
\label{table:perturbations}
\end{center}
\end{table}
%%%%%%%%%%%%%%%%%%% Table %%%%%%%%%%%%%%%%%%%%%%%%%%%%%%%%%%%%%

%%%%%%%%%%%%%%%%%%%%%%%%%%%%%%%%%%%%%%%%%%%%%%%%%%%%%%%%%%%%%%%%%%
\subsection{Third-order perturbation $X_{1,1}$}
\label{sec:third} 
%%%%%%%%%%%%%%%%%%%%%%%%%%%%%%%%%%%%%%%%%%%%%%%%%%%%%%%%%%%%%%%%%%

As we see later, the second-order perturbations are not enough to determine the difference of the entropy between a non-uniform black brane and a uniform one of the same mass. 
Knowledge is required of the correction to the critical wave number, $k_{1}$, which appears at this third order. 
At the third order $O(\epsilon^3)$, new KK modes $X_{1, 1}$ and $X_{3, 0}$ appear independently. 
The equations of the latter modes do not contain  $k_{1}$, and so we are not interested in these modes. 
The KK modes $X_{1, 1}$ are the higher-order counterparts of $X_{1,0}$; the form of the homogeneous equation is identical to Eq. (\ref{eq:first-master}). 
Then the master equation for $b_{1,1}$ takes the form of
\begin{eqnarray}
 {\mathcal L}_1 b_{1,1}
 &=& - \frac{2k_0 k_1}{f}  b_{1,0} + S_{b_{1,1}}(b_{0, 1}, b_{1, 0}, b_{2, 0}; k_0).
 \label{eq:b_{1, 1}}
\end{eqnarray}
The function $a_{1, 1}(y)$ can be solved algebraically as before. 
The source term $S_{b_{1,1}}$ is independent of $k_1$, and the algebraic equation for $a_{1,1}$ does not explicitly contain $k_1$. 
The dependence of $a_{1,1}$ on the critical wave number $k_1$ is only through $b_{1,1}$. 
The horizon boundary condition is given by
\begin{multline}
b_{1,1}' + \frac{2d-6-k_0^2}{d-3} b_{1,1} 
= 
 \frac{2k_1 k_0 }{d-3} b_{1, 0}
 + \frac{b_{1, 0}}{8(d-3)}\\
\times\left\{
 -2(d-3)(8 a_{0, 1} - 8b_{0, 1} + b_{1, 0}^2 - 4 b_{2,0}) 
 + (d-2)k^2_0 \left[ 8b_{0, 1} + (d-2)b_{1, 0}^2 + 4b_{2,0}\right]
\right\},
\end{multline}
where all functions are evaluated at $y=1$.
We can solve this second-order equation in the same manner used for the first-order perturbations. The main difference is that the present case is a two-parameter shooting problem; $b_{1,1} (1)$ is not arbitrary. 
A regular solution exists only for an appropriate set of $b_{1,1}(1)$ and $k_1$. 
We have performed the two-parameter shooting, and the results for various dimensions are summarized in Table \ref{table:k0} and \ref{table:perturbations} (see also Fig. \ref{fig:perturbations}). 
We should notice that the sign of $k_1$ changes above $\mathcal D = 13$:
\begin{eqnarray}
 k_1 > 0 \quad (\mathcal D \leq 13),\nonumber
 \\
 k_1 < 0 \quad (\mathcal D >13).
\end{eqnarray}
This would be a precursor to the existence of a critical dimension.

%%%%%%%%%%%%%%%%%%%%%%%%%%%%%%%%%%%%%%%%%%%%%%%%%%%%%%%%%%%%%%%%%%
\section{Thermodynamics}
\label{sec:thermodynamics}
%%%%%%%%%%%%%%%%%%%%%%%%%%%%%%%%%%%%%%%%%%%%%%%%%%%%%%%%%%%%%%%%%%

We are now ready to calculate the thermodynamical quantities of the perturbative non-uniform solutions. 
From Eq.~(\ref{eq:M,T,S,Q}), we obtain the mass $M$, temperature $T$, entropy $S$ and charge $Q$ for a non-uniform smeared black brane. 
The differences for these quantities between the non-uniform and the uniform critical solutions are  
\begin{eqnarray}
&&     \frac{\delta M}{M} 
    = -\frac{\chi_{1}  \epsilon^2}{(d-2)/(d-3) + \sinh^2 \alpha} +O(\epsilon^4),
\nonumber
\\
&&\frac{\delta T}{T}=\frac{1}{\sqrt{A_{h}}}-1=-\frac{1}{2}a_{0, 1}(1)\epsilon^2+O(\epsilon^4),
\nonumber
\\
&&\frac{\delta S}{S}=\sqrt{A_{h}}-1=\frac{1}{2}a_{0, 1}(1)\epsilon^2+O(\epsilon^4),\nonumber
\\
&&\frac{\delta Q}{Q}=0.
\end{eqnarray}
These physical quantities are expressed in terms of the perturbative quantities, which are calculated without fixing the asymptotic periodicity of the circle. 
It is easy to compare the obtained physical quantities with others if they are given in a frame where the asymptotic length of the circle is fixed.
Without transforming physical quantities into such a frame, we introduce variables that are invariant under rigid rescalings of the entire solution.
Once physical quantities are expressed in terms of the invariant variables, we do not need to care about the variation of the asymptotic periodicity.

The invariant quantities can be obtained by multiplying $K$ by suitable powers:
\begin{eqnarray}
&&  \frac{\delta \mu }{\mu}\equiv\frac{\delta M }{M } 
    + (d-3) \frac{\delta K}{K} 
    = \mu_{1}\epsilon^{2}+O(\epsilon^4) ,\nonumber
\\
  &&  \frac{\delta \tau }{\tau } \equiv
    \frac{\delta T }{T } -  \frac{\delta K}{K} 
    =
    \tau_{1}\epsilon^{2}+O(\epsilon^4) ,\nonumber
\\
&&  \frac{\delta s}{s } \equiv
    \frac{\delta S }{S } + (d-2) \frac{\delta K}{K} 
    =  s_{1}\epsilon^{2}+O(\epsilon^4),\nonumber
\\
&&  \frac{\delta q}{q}  \equiv \frac{\delta Q}{Q}+(d-3)\frac{\delta K}{K} 
    = q_{1}\epsilon^2+O(\epsilon^4),
    \label{eq:rescaling invariants}
\end{eqnarray}
where the leading-order coefficients are given by
\begin{eqnarray}
&&\mu_{1}=-\frac{\chi_{1}}{(d-2)/(d-3)+\sinh^2\alpha}+(d-3)\frac{k_{1}}{k_{0}},
\nonumber
\\
&& \tau_{1}=-\frac{1}{2}a_{0, 1}(1)-\frac{k_{1}}{k_{0}},
\nonumber
\\
&&s_{1}=\frac{1}{2}a_{0, 1}(1)+(d-2)\frac{k_{1}}{k_{0}},
\nonumber
\\
&&
q_{1}=(d-3)\frac{k_{1}}{k_{0}}.
\end{eqnarray}
Here we see the reason why we have integrated the perturbations up to order $O(\epsilon^3)$. 
The correction to the wave number $k_{1}$ is necessary to obtaining invariant quantities.

Note that non-extremal solutions obey the first law,
\begin{eqnarray}
   dM (R_0, \alpha)
 = T(R_0, \alpha) dS(R_0, \alpha)  + \Phi (\alpha) dQ(R_0, \alpha),
\label{eq:first-law with Q}
\end{eqnarray}
provided the asymptotic circle length is fixed. 
We can confirm this directly from 
Eq. (\ref{eq:M,T,S,Q}) by utilizing the first law for the uncharged system, which gives
\begin{eqnarray}
    \frac{1}{2A_h}\frac{ d  A_h }{d R_0}
    = - \frac{ d-3}{R_0} \chi - \frac{ d \chi }{d R_0}. 
\label{eq:first-law A_h}
\end{eqnarray}

%%-----------------------------------------------
\subsection{Vacuum black branes}
\label{sec:neutral}
%%-----------------------------------------------

To begin, let us first consider the thermodynamics of neutral black branes for vacuum spacetime.\footnote{
Since there is no charge in this system, the metric (\ref{eq:metric}) is identical to that of the (neutral vacuum) black $(p+1)$-brane.
}
What we are interested in is the difference between the entropy of a uniform brane and that of a non-uniform one of the same mass.
The difference is evaluated as follows:
\begin{eqnarray}
&& \frac{ S_{\mathrm{NU}} - S_{\mathrm{U}} }{ S_{\mathrm{U}}}
  = \sigma_{1}\epsilon^{2}+\sigma_{2}\epsilon^{4}+O(\epsilon^{6}),
\nonumber
\\
&& 
\sigma_{1}=\mu_{1}-\frac{d-3}{d-2}s_{1},
\quad
\sigma_{2}=-\frac{d-2}{2(d-3)}\left(\tau_{1}+\frac{1}{d-3}\mu_{1}\right)\mu_{1}, 
\label{eq:sigma-neutral}
\end{eqnarray}
where we have used the Smarr formula $TS/M=(d-2-n)/(d-1)$ and the tension of uniform strings, $n=1/(d-2)$. 
This equation can be easily derived by starting from the frame with $\delta K=0$ and re-expressing the results in terms of the invariant quantities (\ref{eq:rescaling invariants}). See Appendix~\ref{sec:appendix1} for more details of the calculation.

As discussed above, the relation (\ref{eq:chi_1 and a_01}) holds for $\chi_1$ and $a_{1, 0}(1)$ from the first law, and then $\sigma_{1}$ vanishes. As a result, the entropy difference arises at $O(\epsilon^4)$. 
The correction $\sigma_{2}$ becomes $\sigma_{2}=  a_{0,1} \mu_1 /4$. Since the numerical value of $a_{0,1}$ is negative (Table~\ref{table:perturbations}), the sign of $\sigma_{2}$ is determined by $\mu_1$. 
Our numerical results for ${\mathcal D} = 4 \sim 15$ are summarized in Table~\ref{table:perturbations}, and they are consistent with previously obtained results~\cite{Gubser:2001ac,Wiseman:2002zc,Sorkin:2004qq}; 
$\sigma_{2}$ is negative for ${\mathcal D} \le 13$, hence the non-uniform phase is not entropically favorable and it implies a first-order phase transition. 
For $ {\mathcal D} > 13$, $\sigma_{2}$ becomes positive, implying a higher-order phase transition as observed in Ref.~\cite{Sorkin:2004qq}.  These are summarized as 
\begin{eqnarray}
  \sigma_{2}   < 0  \quad  ({\mathcal D} \le 13),
\nonumber
\\
  \sigma_{2}  >0  \quad  ({\mathcal D} > 13).
\label{eq:simga2}
\end{eqnarray}

The vanishing of $\sigma_{1}$ can be used as a check of numerics. In fact, our numerical results give vanishingly small $\sigma_{1}$ for relatively low dimensions (${\mathcal{D}} \lesssim 9$).  For much higher dimensions, however, it is hard to read off $\chi_{1}$ numerically because of the rapid decay of the zero modes.  
Thus, we have calculated $\chi_{1}$ by using the relation (\ref{eq:chi_1 and a_01}), which allows very precise determination of physical quantities because $a_{0,1}$ is a local quantity. 
Moreover, the results obtained here are very simple and concise compared to those obtained by the general perturbation theory~\cite{Gubser:2001ac}; the invariant quantities (\ref{eq:rescaling invariants}) under rigid rescalings are given only by 3 variables, i.e., $a_{0,1}, k_0$, and $k_1$, whereas for the general perturbation theory many quantities enter into the estimation of the scheme independent quantities.

Next, we consider a canonical ensemble, i.e., fixed temperature. Helmholtz free energy $F =M -T S $ for the neutral solutions is given by
\begin{eqnarray}
    F =
    \mathcal{G}
   \left(\frac{LR_{0}}{2\pi}\right)^{d-3}
    \left[1-(d-3)\chi  \right].
\label{eq:F0}
\end{eqnarray}
The free energy of the uniform strings is 
$ F_U = {\mathcal G}\left[ { (d-3) }/{4\pi T}\, \right]^{d-3} $.
The difference of the free energy between the uniform and the non-uniform solutions for the same temperature is evaluated as  
\begin{eqnarray}
   \frac{F_{\rm NU}-F_{\rm U}}{F_{\rm U}}
= 
   \rho_{2} \epsilon^4 + O (\epsilon^6),  
\qquad
    \rho_{2}  = -\frac{1}{2}(d-3)(d-2) 
    \left( \tau_{1}+ \frac{1}{d-3}\mu_{1} \right) \tau_{1}.
    \label{eq:rho2}
\end{eqnarray}
See Appendix \ref{sec:appendix1} for the derivation of Eq.~(\ref{eq:rho2}).
The structure of the coefficients $\rho_2$ is quite similar to that in the microcanonical ensemble (\ref{eq:sigma-neutral}), except the overall coefficient, which is $\tau_1$ in the present case. 
It is interesting that the invariant variable $\tau_1$ changes its sign at the different dimension $\mathcal D$ from that for $\mu_1$.
$\tau_1$ is negative  for ${\mathcal D} \le 12$ and positive for ${\mathcal D} > 12$ (Table~\ref{table:perturbations}).
Consequently, the free energy of the non-uniform phase becomes favorable at ${\mathcal D} > 12$;
\begin{eqnarray}
  \rho_{2}   > 0  \quad  ({\mathcal D} \le 12),
\nonumber
\\
  \rho_{2}  < 0  \quad  ({\mathcal D} > 12).
\label{eq:rho2-result}
\end{eqnarray}
It is remarkable that $\mathcal{D}=13$, which is the critical dimension in the microcanonical ensemble, is not special in the canonical ensemble.

%%--------------------------------------------------------
\subsection{Weakly charged smeared black branes}
%%--------------------------------------------------------

General thermodynamic properties of the charged (dilatonic) black branes are rather complicated compared with the vacuum cases. 
Besides, the general formula that is given in terms of the invariant quantities is untractable, so we focus attention on two limiting cases: a weakly charged case and a near-extremal case. The latter will be discussed in the next subsection.

Firstly, we consider weakly charged smeared black branes in a microcanonical ensemble (fixed mass and charge).
Expanding the entropy difference for small charge $Q\ll M $, we find
\begin{eqnarray}
&&\frac{S_{\rm NU}-S_{\rm U}}{S_{\rm U}}
\approx \left[\sigma_{2} + \left( \frac{Q}{M_{\chi=0}}\right)^2
\delta \sigma_2  +O(Q^4)\right]~ \epsilon^4,
\nonumber
\\
&&\delta\sigma_{2}=\frac{(d-2)^2}{2(d-3)^2}\left\{\left[\tau_{1}+q_{1}-\frac{2(d-4)}{d-3}\mu_{1}\right]q_{1}+\left(\frac{d-5}{d-3}\mu_{1}-\tau_{1}\right)\mu_{1}\right\},
\label{eq:entropy-diff}
\end{eqnarray}
where $\mu_1$ and $\tau_1$ are evaluated at ${{\alpha=0}}$,  and we note that $Q/M$ is invariant under rigid rescalings. In Appendix~\ref{sec:appendix1}, we will show how Eq.~(\ref{eq:entropy-diff}) is derived.
As in the vacuum case, the leading entropy difference appears at $O(\epsilon^4)$, and no correction appears at a lower order.
This is not due to an accidental cancellation under the approximation of weak charge, but we can show after a tedious calculation that leading corrections are always $O(\epsilon^4)$ for general charge $Q$. 
The correction term $\delta \sigma_2$ is positive with numerical values of $O(1)$ for ${\mathcal D} \le 15$, whereas $\delta \sigma_2$ becomes negative for a very large value of $\mathcal D$.
 (We have confirmed this up to $\mathcal{D}=20$.)
In any case, the new correction is sufficiently small compared to $\sigma_2$, as long as the assumption of weak change is valid, and hence the charge and dilaton do not change the thermodynamic phase structure.  
One might think that the phase structure is drastically affected by the bulk fields with an arbitrary charge, because the correction $ \delta \sigma_2$ counteracts the leading term $\sigma_2$ in the above case.
As we will see below, however, the same result of (\ref{eq:simga2}) holds even for the near-extremal limit in a microcanonical ensemble.

\begin{figure}[tbp]  %%[thbp]
\begin{center}
\includegraphics[width=9cm]{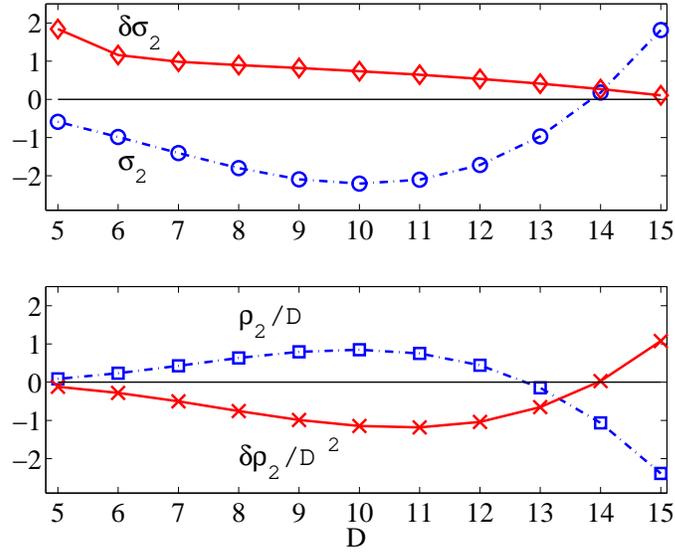}
\caption{
\label{fig:sigma2etc}
Dependence of the coefficients on the effective spacetime dimension ${\mathcal D}$. $\rho_2$ and $\delta \rho_2$ are divided by $\mathcal{D}$ and ${\mathcal D}^2$, respectively.
The sign change of $\sigma_2$ is at ${\mathcal D} > 13$, whereas the sign change of $\rho_2$ is at ${\mathcal D} > 12$.  
}
\end{center}
\end{figure}

Next, we turn to a canonical ensemble for weakly charged branes.
The difference of the free energy for the same temperature and charge is 
\begin{eqnarray}
&&  \frac{F_{NU} - F_U}{F_U} 
  \approx
  \left[
     \rho_2 +   \left( \frac{Q}{M_{\chi=0}}\right)^2 \delta\rho_2
  \right]  \epsilon^4, 
\nonumber
\\
  && \delta\rho_{2} = 
     \frac{(d-2)^2}{2(d-3)} 
     \Bigl\{ (d-3)(s_{1}+\tau_{1})\tau_{1}-\left[q_{1}+2(d-3)\tau_{1}\right]q_{1}
     \Bigr\}  .
     \label{eq:free-energy-diff}
\end{eqnarray}
See Appendix \ref{sec:appendix1} for the derivation of Eq.~(\ref{eq:free-energy-diff}).
The dimensional dependence of $\delta\rho_2$ is depicted in Fig. \ref{fig:sigma2etc}. 
From the figure we see that $\delta\rho_2$ is negative for $\mathcal D \le13$, hence, the correction counteracts $\rho_2$. 
The correction in this weakly charged case is too small to change the phase structure; interestingly, the effect becomes evident near the extremality, as we will see in the next section.

%%%%%%%%%%%%%%%%%%%%%%%  Fig.  %%%%%%%%%%%%%%%%%%%%%%%%%%%%%%%
%%%%%%%%%%%%%%%%%%%%%%%%%%%%%%%%%%%%%%%%%%%%%%%%%%%%%%%%%%%%%%

%%-----------------------------------------------
\subsection{Near-extremal smeared black branes}
%%-----------------------------------------------

Let us finally consider the thermodynamic behavior near the extremal state. 
A near-extremal limit can be realized by taking $\alpha$ to infinity.
We define energy above extremality $\mathcal{M}$, ``reduced temperature'' $\mathcal{T}$ and ``reduced entropy'' $\mathcal{S}$ as follows:
\begin{eqnarray}
&&
    \mathcal{T} \equiv 
    \lim_{\alpha\to\infty}\sqrt{Q}T 
    = \left[\frac{(d-3)^{3} {\mathcal G}  }{ 4 A_{h}L^2R_{0}^{2}  }
     \left(\frac{L R_{0}}{2\pi}\right)^{d-3}\right]^{1/2},
\nonumber
\\
&&
    \mathcal{S} \equiv \lim_{ \alpha\to\infty}\frac{S}{\sqrt{Q}}
    = 
    \left[ \frac{16\pi^2 A_{h} {\mathcal G} }{(d-3) }
    \left(\frac{L R_{0}}{2\pi}\right)^{d-1}\right]^{1/2},
\nonumber
\\
&&
    \mathcal{M} \equiv \lim_{\alpha\to\infty}(M-Q)
    = 
    {\mathcal G} \left(\frac{L R_{0}}{2\pi}\right)^{d-3}
    \left[ \frac{d-1}{2}-(d-3) \chi \right],
\nonumber
\\
&&
    {\mathcal M}_{\rm el}  \equiv \lim_{\alpha\to\infty}(M_{el}-Q)
    = 
    -   {\mathcal G}  \left(\frac{L R_{0}}{2\pi}\right)^{d-3}
 \frac{d-3}{2}.
\label{eq:NE MTS}
\end{eqnarray}
Note that $\mathcal{M} - {\mathcal{M}}_{\rm el}=M|_{\alpha=0}$ and ${\mathcal{T S}}= T S|_{\alpha=0}$ hold.
Using Eq. (\ref{eq:first-law A_h}), we can rewrite the first law in terms of the reduced quantities near the extremal state: 
\begin{eqnarray}
    d\mathcal{M}(R_0)=\mathcal{T}(R_0)d\mathcal{S}(R_0).
    \label{eq:first-law-extremal}
\end{eqnarray}
In addition, we notice that the Smarr formula also takes the same form as that for the uncharged case.

The invariant fractional changes of the reduced quantities are given by
\begin{eqnarray}
&&
\frac{\delta \bar{\mu}}{\bar{\mu}}\equiv\frac{\delta \mathcal{M}}{\mathcal{M}}+(d-3)\frac{\delta K}{K} 
= \bar{\mu}_{1}\epsilon^2+O(\epsilon^4),\nonumber
\\
&&\frac{\delta \bar{\tau}}{\bar{\tau}}
\equiv\frac{\delta \mathcal{T}}{\mathcal{T}}+\frac{1}{2}(d-5)\frac{\delta K}{K}
= \bar{\tau}_{1}\epsilon^2+O(\epsilon^4),\nonumber
\\
&&\frac{\delta \bar{s}}{\bar{s}}
\equiv\frac{\delta \mathcal{S}}{\mathcal{S}}+\frac{1}{2}(d-1)\frac{\delta K}{K}
 = \bar{s}_{1}\epsilon^2+O(\epsilon^4),
\end{eqnarray}
where the leading-order coefficients are given by
\begin{eqnarray}
\bar{\mu}_{1} &=&-\frac{2(d-3)}{d-1}\chi_{1}+(d-3)\frac{k_{1}}{k_{0}},\nonumber
\\
\bar{\tau}_{1}&=&-\frac{1}{2}a_{0, 1}(1)+\frac{1}{2}(d-5)\frac{k_{1}}{k_{0}},\nonumber
\\
\bar{s}_{1} &=& \frac{1}{2}a_{0, 1}(1)+\frac{1}{2}(d-1)\frac{k_{1}}{k_{0}}.
\end{eqnarray}

We consider first the thermodynamics in a microcanonical ensemble (fixed energy $\mathcal{M}$). The entropy difference between the uniform and non-uniform solutions for the same energy $\mathcal{M}$ is computed with the first law (\ref{eq:first-law-extremal}) as
\begin{eqnarray}
&&  \frac{\mathcal{S}_{\rm NU}-\mathcal{S}_{\rm U}}{\mathcal{S}_{\rm U}}
   = \bar{\sigma}_{2}\epsilon^4+O(\epsilon^6), \;\;\;
    \bar{\sigma}_{2} 
    = \frac{d-1}{4(d-3)}\left[\frac{d-5}{2(d-3)}\bar{\mu}_{1}-\bar{\tau}_{1}\right]\bar{\mu}_{1}.
\end{eqnarray}
The numerical coefficients of $\bar{\mu}_{1}$ and $\bar{\tau}_{1}$ differ from those of (\ref{eq:sigma-neutral}), due to the ``corrections" ${\mathcal M}_{\mathrm{el}}$ to the mass in the Smarr formula (\ref{eq:smarr}), i.e., $\mathcal{T}\mathcal{S}/\left(\mathcal{M}-\mathcal{M}_{\mathrm{el}}\right)=2(d-3) /(d-1)$ for the uniform phase ($\chi=0$). 
The entropy difference again arises at $O(\epsilon^4)$ as the neutral case.
As we can see from Table~\ref{table:near-extremal}, the entropy of the non-uniform phase is less than the entropy of the uniform phase for a given energy $\mathcal{M}$ for $ {\mathcal{D}} \le 13$. In other words, the non-uniform phase is disfavored thermodynamically over the uniform phase.
For ${\mathcal{D}} > 13$, on the other hand, the non-uniform phase is favorable over the uniform phase: 
\begin{eqnarray}
 \bar{\sigma}_{2} < 0 \quad ({\mathcal{D}} \le 13),
\nonumber
\\
 \bar{\sigma}_{2} > 0 \quad ({\mathcal{D}} > 13).
\end{eqnarray}
The critical dimension at which the sign of $\bar{\sigma}_{2}$ changes is same as the one for the uncharged case.

%%%%%%%%%%%%%%%%%%% Table %%%%%%%%%%%%%%%%%%%%%%%%%%%%%%%%%%%%%
\begin{table}[tb]
\begin{center}
\begin{tabular}{rc||c|c|c|c|c|c|c|c|c|c|c }
%%%%%%%%%%%%%%%%%%%%%%%%%
\hline\hline
&$\mathcal{D}$ & 5 & 6 & 7 & 8 & 9 & 10 & 11 & 12 & 13 & 14 & 15
\\ \hline
 & $\bar{\sigma}_{2}$&-0.53 &-0.91 & -1.32 & -1.71 &-2.00 &-2.10& -2.00& -1.6 & -0.9 &~0.3&~1.9
\\ \hline
& $\bar{\rho}_{2}$ &~0.18&---&-15.7&-10.4&-9.19&-8.40&-7.43&-6.0&-3.8&
 -0.8&~3.3
\\
\hline \hline 
\end{tabular}
\caption[short]{
Numerical results of invariant quantities for near-extremal smeared black branes. 
}
\label{table:near-extremal}
\end{center}
\end{table}
%%%%%%%%%%%%%%%%%%% Table %%%%%%%%%%%%%%%%%%%%%%%%%%%%%%%%%%%%%

We now turn to a canonical ensemble (fixed temperature $\mathcal{T}$).
We define the ``free energy'' of the solution as
\begin{eqnarray}
    \mathcal{F} \equiv \mathcal {M}- \mathcal{T} \mathcal{S}.
\end{eqnarray}
The free energy of the uniform solution is 
\begin{eqnarray}
    \mathcal{F}_{U} (R_0)= - \frac{ d-5}{d-1}  {\mathcal M}(R_0), 
\end{eqnarray}
which vanishes for $d=5$ and is negative (positive) for $d>5$ ($d<5$).
The difference of free energy between the uniform and non-uniform solutions for the same temperature is given by
\begin{eqnarray}
\frac{\mathcal{F}_{\rm NU}-\mathcal{F}_{\rm U}}{\mathcal{F}_{\rm U}}
=
\bar{\rho}_{2}\epsilon^4 + O(\epsilon^6), \;\;\;\;
  \bar{\rho}_{2} = 
  \frac{d-1}{d-5}  
  \left(\frac{1}{2}\bar{\mu}_{1} - \frac{d-3}{d-5}\bar{\tau}_{1}\right)\bar{\tau}_{1},
\end{eqnarray}
where ${\mathcal D} \neq 6$ ($d\neq 5$). The exceptional case of ${\mathcal D}=6$ has to be treated separately because the free energy of the uniform solution vanishes. This case will be discussed elsewhere (see Ref.~\cite{Harmark:2003dg}).  
Table \ref{table:near-extremal} summarizes our results. 
We see that $\bar{\rho}_2$ is positive for $ {\mathcal D } = 5$ and $ {\mathcal D } > 14$. 
This change of sign at the lower dimensions is in contrast to the previously studied cases in which the change of sign occurs only at relatively higher dimensions. 
However, we should recall that the free energy $\mathcal{F}_{\rm U}$ of the uniform solution changes its sign depending on the effective dimension. 
Consequently, the non-uniform phase becomes thermodynamically favorable over the uniform phase only at ${\mathcal D} > 14$: 
\begin{eqnarray}
    {\mathcal F}_{\rm U} <  {\mathcal F}_{\rm NU} \quad  ({\mathcal D} \le 14),\nonumber
\\
    {\mathcal F}_{\rm U} >  {\mathcal F}_{\rm NU} \quad  ({\mathcal D} > 14).
\end{eqnarray}
This result differs from the vacuum case (\ref{eq:rho2-result}). 
We have observed in the weakly charged case that the correction due to the charge has a tendency to change the phase structure. 
The large correction near the extremality shifts the critical dimension upward, although the correction cannot change the phase structure significantly.

%%============================================
\section{Conclusion}
\label{sec:conclusion}
%%============================================

We have investigated the uniform and non-uniform black branes smeared on a transverse circle by constructing perturbative solutions explicitly.
We have made use of the HO metric ansatz in our perturbation analysis.
In the original perturbation scheme \cite{Gubser:2001ac} based on the general metric ansatz,
analysis requires a good deal of subtle maneuvering.
Making use of the HO metric has made the analysis enormously simple and clear in many respects. 
First of all, the EOMs of the charged dilatonic black branes are the same as those of the vacuum black branes on a circle. Consequently, we do not need to perturb the dilaton and form fields at any order, and the task is reduced to solving only one master equation at each order. Owing to such simplicity, the asymptotic behavior of the perturbation and the (analytically approximate) GL critical wave number are easily obtained. Besides, all physical quantities are characterized by a single parameter $a_{0,1}$.

The existence of a static mode means an onset of unstable modes, hence, the smeared black branes suffer from the GL instability irrespective of their charge. However, this does not mean that they provide a counterexample of the GM conjecture since there is no parameter region in which local thermodynamic stability is realized, except at the extremality.

Having performed the perturbation up to the third order, we obtain the phase structure around the critical uniform solution. 
We have begun with studying vacuum black branes. 
In the microcanonical ensemble, the thermodynamic system has the critical dimension ${\mathcal D_*} = D-p=13$, at which the nature of phase transition between the uniform and non-uniform branes changes \cite{Sorkin:2004qq}.
On the other hand, we found that the critical dimension changes if we change the ensemble.
In the canonical ensemble, the critical dimension is ${\mathcal D_*}=12$. 
Furthermore, if the system is near the extremal state, the critical dimension in the canonical ensemble becomes ${\mathcal D_*}=14$, whereas the critical dimension in the microcanonical ensemble does not change. 
We think that a specific value for a critical dimension is not universal. 
It depends on matters in the bulk and type of ensemble.

While the third-order perturbations have been performed to obtain meaningful leading-order corrections, the next-order corrections are also interesting to study. 
To obtain such corrections, we need to perform the perturbations up to the fifth order, and it would appear possible to obtain them. 
We think that the perturbative results in this paper give a good theoretical basis for understanding and constructing fully nonlinear solutions of the HO ansatz.
Fully nonlinear solutions will allow us to explore the thermodynamic phase structure for an arbitrary charge, which might have significant consequences.
We will discuss the fully nonlinear (numerical) solutions in a forthcoming paper. 
It is clearly interesting if a critical dimension appears at much lower dimensions, because there is no sensible quantum theory of gravity for $D > 10,11$. 
Such a situation might be possible if we consider other types of bulk fields or configurations.

%%%%%%%%%%%%%%%%%%%%%%%%%%%%%%%%%%%%%%%%%%%%%%%%%%%%%%%%%%%%
\acknowledgments
 H. K. is supported by the JSPS. U. M. is partially supported by a Grant for The 21st Century COE Program (Holistic Research and Education Center for Physics 
Self-Organization Systems) at Waseda University.
%%%%%%%%%%%%%%%%%%%%%%%%%%%%%%%%%%%%%%%%%%%%%%%%%%%%%%%%%%%%%%%%%%%%%%%%

%%%%%%%%%%%%%%%%%%%%%%%%%%%%%%%%%%%%%%%%%%%%%%%%%%%%%%%%%%%%%%%%%%
%bibliographystyle{plain} 
%\bibliographystyle{apsrev} 
%%\bibliography{RSBHs.bib}
%%\input{RSBHs_bbl.tex} 
%%%%%%%%%%%%%%%%%%%%%%%%%%%%%%%%%%%%%%%%%%%%%%%%%%%%%%%%%%%%

\appendix
%%============================================
\section{Vacuum Einstein equations}
\label{sec:vacuum}
%%============================================

In this appendix we give explicit representation of the Einstein equations. 
From the ansatz (\ref{eq:metric}), the metric of the ($d+1$)-dimensional vacuum black branes on a circle is 
\begin{eqnarray}
    ds^2_{d+1} 
    = -f dt^2 + \left( \frac{L}{2\pi}\right)^2
    \left[
     \frac{e^a }{f} dR^2 + e^{a -(d-2)b} dv^2 + e^{b}R^2 d\Omega_{d-2}^2 
    \right],
\end{eqnarray}
where we have taken $A(R,v)=e^{a}$ and $B(R,v)=e^{b}$. 
The nontrivial components of the vacuum Einstein equations ${\mathcal{R}}^i_j =0$ 
are given by
\begin{eqnarray}
{\mathcal R}^v_v&=& \frac{e^{-a}}{2} \left( \frac{2\pi}{L}\right)^2
    \biggl(
     f \left[(d-2)b''-a''\right]-e^{(d-2)b} \left[ (d-2) \ddot{b}+\ddot{a}
    \right]
\cr
%\nonumber \\ 
&&\hspace{0.95cm}
    + \frac{1}{2R} 
    \left\{2(d-3+f)\left[(d-2)b' - a'\right] - e^{(d-2)b} R (d-2) (d-1) {\dot{b}}^2
    \right\}
    \biggr) ,
    \cr
%\nonumber\\
    {\mathcal R}^R_R
&=& - \frac{e^{-a}}{2} \left( \frac{2\pi}{L}\right)^2
    \biggl(
    f a'' + \ddot{a} \,e^{(d-2)b}
\cr
%\nonumber\\
&&\hspace{1.35cm}     +  (d-2) \dot{a} \dot{b} \, e^{(d-2)b} 
  +  (d-2)f  
     \left\{
   \left[ \frac{2}{R} - a'+ \frac{(d-1)b'}{2}  \right] b'    - \frac{a'}{R} 
     \right\}
    \biggr),
\cr
%\nonumber\\
    {\mathcal R}^\theta_\theta
&=& -  \frac{e^{-a}}{2} \left( \frac{2\pi}{L}\right)^2
    \biggl[
     \frac{2(d-3)}{R^2}(1-e^{a-b})
%\cr
\nonumber\\
&&\hspace{3cm}   + \frac{1}{R}(d-3 +f) b'
    + e^{(d-2)b} (d-2) {\dot{b}}^2
    + f b'' + e^{(d-2)b} \ddot{b}
    \biggr],
\cr
%\nonumber\\
    {\mathcal R}^R_v
&=&   \frac{e^{-a}}{4} \left( \frac{2\pi}{L}\right)^2
    \biggl(
     \frac{1}{R}
     \left\{ d-3 + f [ d-1 + (d-2)R b' \,]\right\} \, \dot{a}
\cr     
%\nonumber\\
&&  \hspace{3.8cm}
   + (d-2) f \left[ a' - \frac{2}{R} - (d-1) b' \right] \dot{b}
 -2 f (d-2) \dot{b}'
    \biggr),
\label{eq:vacuumEinstein}
\end{eqnarray}
where the prime and dot stand for the derivatives with respect to $R$ and $v$, respectively. 
Since the angular component ${\mathcal R}^\theta_\theta$ does not contain any derivatives of $a(R,v)$, $a(R,v)$ can be given in terms of $b(R,v)$ and its derivatives.

%%%%%%%%%%%%%%%%%%%%%%%%%%%%%%%%%%%%%%%%%%%%%%%%%%%%%%%%%%%%%%%%%%%%%%%%
\section{Entropy and free energy difference}
\label{sec:appendix1}
%%%%%%%%%%%%%%%%%%%%%%%%%%%%%%%%%%%%%%%%%%%%%%%%%%%%%%%%%%%%%%%%%%%%%%%%

In this appendix, we derive the formulae of the entropy difference and free energy difference, i.e., Eqs. (\ref{eq:sigma-neutral}), (\ref{eq:rho2}), (\ref{eq:entropy-diff}) and (\ref{eq:free-energy-diff}).

First, we expand the thermodynamical quantities by $\epsilon$ as follows:
\begin{eqnarray}
M=\sum_{p=0}^{\infty}M_{p}\epsilon^{2p},\;\;\;T=\sum_{p=0}^{\infty}T_{p}\epsilon^{2p},\;\;\;
S=\sum_{p=0}^{\infty}S_{p}\epsilon^{2p},\;\;\;Q=\sum_{p=0}^{\infty}Q_{p}\epsilon^{2p}.
\end{eqnarray}
The entropy difference between the uniform and non-uniform branes of the same mass and charge is expressed as
\begin{eqnarray}
\frac{S_{\rm NU}-S_{\rm U}}{S_{\rm U}}
&=&\frac{\sum_{p=0}^{\infty}S_{p}\epsilon^{2p}}{S_{0}+\Delta S_{0}} - 1
\nonumber\\
&=&\frac{S_{1}}{S_{0}}\epsilon^2+\frac{S_{2}}{S_{0}}\epsilon^4-\frac{S_{1}}{S_{0}}\frac{\Delta S_{0}}{S_{0}}\epsilon^2-\frac{\Delta S_{0}}{S_{0}}+\left(\frac{\Delta S_{0}}{S_{0}}\right)^2+O\left(\epsilon^6\right),
\label{entropy-diff}
\end{eqnarray}
where $\Delta S_{0}$ is the entropy change of a \textit{uniform} brane due to the change of the mass and charge, which we denote by $\Delta M_{0}$ and $\Delta Q_{0}$, respectively.

Let us focus on the quantities $S_{1}/S_{0}$ and $S_{2}/S_{0}$ in Eq.~(\ref{entropy-diff}). From the first law with a fixed scale of the circle ($\delta L=0$), we have
\begin{eqnarray}
M_{1}=T_{0}S_{1}+Q_{1}\tanh\alpha,\;\;\;M_{2}=T_{0}S_{2}+\frac{1}{2}T_{1}S_{1}+Q_{2}\tanh\alpha.
\label{eq:1stlaw1}
\end{eqnarray}
From Eq.~(\ref{eq:M,T,S,Q}), the mass of a uniform brane is written in two ways:
\begin{eqnarray}
M_{0}=\left(\frac{d-2}{d-3}+\sinh^2\alpha\right)T_{0}S_{0}\;\;\;
\mbox{and}\;\;\;
M_{0}=\left[1+\frac{d-2}{(d-3)\sinh^2\alpha}\right]Q_{0}\tanh\alpha.
\label{eq:uniform-Smarr}
\end{eqnarray}
From these equations (\ref{eq:1stlaw1}) and (\ref{eq:uniform-Smarr}), we can write the entropy difference, $S_{1}/S_{0}$ and $S_{2}/S_{0}$, in terms of $M_{1}/M_{0}$, $M_{2}/M_{0}$, $Q_{1}/Q_{0}$, $Q_{2}/Q_{0}$ and $T_{1}/T_{0}$. 

Next, we focus on the quantity $\Delta S_{0}/S_{0}$ in Eq.~(\ref{entropy-diff}). To express the entropy of uniform brane in terms of mass and charge, we solve Eq.~(\ref{eq:M,T,S,Q}) for $\alpha$ and $R_{0}$ with $\chi=0$,
\begin{eqnarray}
    \sinh ^2\alpha 
    &=& \frac{1}{2 (1-Q_0^2/M_0^2)} 
    \left[
    -1 + 2 \frac{d-2}{d-3} \left( \frac{Q_0}{M_0}\right)^2 
    + \sqrt{
    1 + \frac{4(d-2)}{(d-3)^2} \left( \frac{Q_{0}}{M_{0}}\right)^2 
    }
    \right],\nonumber
\\
     \frac{L R_0}{2\pi}  
    &=& 
    \left[
    \frac{ {\mathcal G} (d-3) }{M_0  }
    \left(
          \frac{d-2}{d-3} + \sinh^2 \alpha
    \right)
    \right]^{-1/(d-3)}.
\end{eqnarray}
Substituting above two relations into the entropy of the uniform brane ($A_{h}=1$) in Eq.~(\ref{eq:M,T,S,Q}), one obtains in the leading order of $Q_{0}/M_{0}$ ($\ll1$)
\begin{eqnarray}
S_{0}(M_{0}, Q_{0})\approx4\pi\mathcal{G}\left[\frac{M_{0}}{(d-2)\mathcal{G}}\right]^{(d-2)/(d-3)}\left[1-\frac{1}{2}\left(\frac{d-2}{d-3}\right)^2\left(\frac{Q_{0}}{M_{0}}\right)^2\right].
\end{eqnarray}
From this expression, we can compute $\Delta S_{0}=S(M_{0}+\Delta M_{0}, S_{0}+\Delta S_{0})-S(M_{0}, S_{0})$.
Taking the mass and charge differences as $\Delta M_{0}\approx M_{1}\epsilon^2+M_{2}\epsilon^4$ and $\Delta Q_{0}\approx Q_{1}\epsilon^2+Q_{2}\epsilon^4$, we obtain $\Delta S_{0}/S_{0}$ in terms of $M_{1}/M_{0}$, $Q_{1}/Q_{0}$ and so on.

Finally, replacing such gauge dependent quantities with gauge independent quantities such as $\mu_{1}$ with $\delta K=0$, we obtain the expressions (\ref{eq:sigma-neutral}) and (\ref{eq:entropy-diff}).

In a similar way, we can derive the formula of the free energy difference, Eqs.~(\ref{eq:rho2}) and (\ref{eq:free-energy-diff}). The free energy difference between uniform and non-uniform branes of the same temperature and charge is given by
\begin{eqnarray}
\frac{F_{\rm NU}-F_{\rm U}}{F_{\rm U}}&=&
    \frac{\sum_{p=0}^{\infty}F_{p}\epsilon^{2p}}{F_{0}+\Delta F_{0}} - 1
\nonumber\\
&=&\frac{F_{1}}{F_{0}}\epsilon^2+\frac{F_{2}}{F_{0}}\epsilon^4-\frac{F_{1}}{F_{0}}\frac{\Delta F_{0}}{F_{0}}\epsilon^2-\frac{\Delta F_{0}}{F_{0}}+\left(\frac{\Delta F_{0}}{F_{0}}\right)^2+O\left(\epsilon^6\right),
\label{free-energy-diff}
\end{eqnarray}
where $\Delta F_{0}$ means the free energy change of a \textit{uniform} brane due to the change of temperature and charge. From the first law with a fixed scale of the circle, we have
\begin{eqnarray}
F_{1}=-S_{0}T_{1}+Q_{1}\tanh\alpha,\;\;\;F_{2}=-S_{0}T_{2}+\frac{1}{2}S_{1}T_{1}+Q_{2}\tanh\alpha.
\label{eq:1st-law-free-energy}
\end{eqnarray}
From Eq.~(\ref{eq:uniform-Smarr}), we can express the free energy of a uniform brane in two ways:
\begin{eqnarray}
F_{0}\approx\left(\frac{1}{d-3}+\alpha^2\right)T_{0}S_{0}\;\;\; \mbox{and} \;\;\; F_{0}\approx\alpha^{-1}\left(\frac{1}{d-3}+\alpha^2\right)Q_{0} \;\;\;(\alpha\ll 1).
\label{eq:eq:uniform-FSTQ}
\end{eqnarray}
 From Eqs.~(\ref{eq:1st-law-free-energy}) and (\ref{eq:eq:uniform-FSTQ}), we can write the free energy difference, $F_{1}/F_{0}$ and $F_{2}/F_{0}$, in terms of $T_{1}/T_{0}$, $T_{2}/T_{0}$, $Q_{1}/Q_{0}$, $Q_{2}/Q_{0}$ and $S_{0}/S_{1}$. 
 
 To calculate $\Delta F_{0}/F_{0}$, we have to express the free energy of uniform branes as a function of temperature and charge. From Eq.~(\ref{eq:M,T,S,Q}), we have
\begin{eqnarray}
\alpha&=&\frac{Q_{0}}{(d-3)\mathcal{G}}\left(\frac{4\pi T_{0}}{d-3}\right)^{d-3}+O\left(Q_{0}^3\right),\nonumber\\
\frac{LR_{0}}{2\pi}&=&\frac{d-3}{4\pi}\left[1-\frac{1}{2}\frac{Q_{0}^2}{(d-3)^2\mathcal{G}^2}\left(\frac{4\pi T_{0}}{d-3}\right)^{2(d-3)}\right]+O\left(Q_{0}^4\right).
\end{eqnarray}
With these relations, the free energy of a uniform brane is given by
\begin{eqnarray}
F_{0}(T_{0}, Q_{0})=\mathcal{G}\left(\frac{d-3}{4\pi T_{0}}\right)^{d-3}\left[1+\frac{Q_{0}^2}{2(d-3)\mathcal{G}^2}\left(\frac{4\pi T_{0}}{d-3}\right)^{2(d-3)}\right].
\end{eqnarray} 
From this, one can compute the free energy difference of a uniform brane as $\Delta F_{0}=F_{0}(T_{0}+\Delta T_{0},Q_{0}+\Delta Q_{0})-F_{0}(T_{0},Q_{0})$. Taking the change of temperature and charge as $\Delta T_{0}\approx T_{1}\epsilon^2+T_{2}\epsilon^4$, $\Delta Q_{0}\approx Q_{1}\epsilon^2+Q_{2}\epsilon^4$, we can write $\Delta F_{0}/F_{0}$ in terms of $S_{1}/S_{0}$, $T_{1}/T_{0}$ and so on. Therefore, replacing these with gauge independent quantities, we obtain Eqs. (\ref{eq:rho2}) and (\ref{eq:free-energy-diff}).

%%%%%%%%%%%%%%%%%%%%%%%%%%%%%%%%%%%%%%%%%%%%%%%%%%%%%%%%%%%%%%%%%%%%%%%%

\end{document}